\documentclass[a4paper,fleqn,usenatbib]{mnras}
\usepackage{txfonts}
\usepackage[T1]{fontenc}
\usepackage{ae,aecompl}

\usepackage{graphicx}

\newcommand{\gamete}{\textsc{GAMETE/QSOdust}}

\newcommand{\msun}{$M_\odot$}

\title[BH seed hosts properties]{Chasing the observational signatures of seed black holes at $z>7$: candidate statistics}

\author[Valiante et al.]{Rosa Valiante$^1$ \thanks{E-mail: rosa.valiante@oa-roma.inaf.it},
Raffaella Schneider$^{1,2}$, Luca Graziani$^1$, Luca Zappacosta$^1$ \\
$^1$ INAF-Osservatorio Astronomico di Roma, via di Frascati 33, 00040, Monteporzio Catone, 
Italy\\
$^2$ Dipartimento di Fisica, Universit\'a di Roma “La Sapienza”, P.le Aldo Moro 2, 00185 Roma, Italy 
}

\date{Accepted . Received }
\pubyear{2017}

\begin{document}
\label{firstpage}
\pagerange{\pageref{firstpage}--\pageref{lastpage}} 
\maketitle

\begin{abstract}
Supermassive black holes (SMBHs) of $10^9-10^{10} \, M_\odot$ were already in place $\sim 13$ Gyr ago, 
at $z>6$.
Super-Eddington growth of low-mass BH seeds ($\sim 100 \, M_\odot$) or less extreme accretion onto 
$\sim 10^5 \, M_\odot$ seeds have been recently considered as the main viable routes to these SMBHs.
Here we study the statistics of these SMBH progenitors at $z\sim 6$. 
The growth of low- and high-mass seeds and their host galaxies are consistently followed using the cosmological data constrained model \textsc{GAMETE/QSOdust}, which reproduces the observed properties of high-$z$ quasars, 
like SDSS J1148+5251.
We show that both seed formation channels can be in action over a similar redshift range, $15<z<18$ and are found in dark matter halos with comparable mass, $\sim 5\times 10^7 \, M_\odot$. However, as long as the systems evolve in isolation (i.e. no mergers occur), noticeable differences in their properties emerge: at $z\geq 10$ galaxies hosting high-mass seeds have smaller stellar mass and metallicity, the BHs accrete gas at higher rates and star formation proceeds less efficiently than in low-mass seeds hosts. At $z<10$ these differences are progressively erased, as the systems experience minor or major mergers and every trace of the BH origin gets lost.
\end{abstract}


\section{Introduction}
The first billion years of the Universe must have been the stage of intense black hole (BH) activity, as shown by the large number of quasars ($>170$) discovered at $z>5.6$ \citep[e.g.][and references therein]{Banados16, Jiang16}.

The luminosity of the brightest ones ($>10^{47} \rm erg/s$) is commonly interpreted as an indication
of  actively accreting supermassive black holes (SMBHs) of $>10^9 \, M_\odot$ in their hosting galaxies 
\citep{Fan01, deRosa11, deRosa14}. To date, the most massive BH observed at $z>6$ is
the one powering the quasar SDSS J0100+2802 \citep{Wu15}, while the most distant SMBH is located 
at $z=7.08$ \citep{Mortlock11}. 
%

In the standard paradigm of BH evolution, a SMBH is expected to be the evolutionary product of a less massive BH seed, which 
subsequently grows via gas accretion and mergers. 
The BH seeds birth mass function 
and the physical mechanisms driving their formation and growth 
are among the most studied issues in the recent literature \citep[see reviews by][]{Volonteri10, Natarajan11, 
LatifFerrara16,JohnsonHaardt16,V17a}.

BHs forming as remnants of Population III (Pop~III) stars have been predicted to have masses 
from few tens to few hundreds (or even thousands) solar masses \citep[e.g.][]{MadauRees01,Heger03,
Yoshida08,Latif13b,Hirano15}. They are also supposed to become supermassive by $z\sim 6-7$ only if their growth is
driven by uninterrupted, although unrealistic, gas accretion at the Eddington rate or, alternatively, 
by short (intermittent) periods of super-critical feeding \citep[][]{Haiman04,Volonteri05,Volonteri06,
TH09,VSD15,Lupi16,P16,P17b}. 

Alternative routes to the first SMBHs are provided by intermediate mass seeds 
($> [10^3-10^4]~M_\odot$) forming in dense stellar cluster either via runaway collisions 
\citep[][]{Omukai08,Devecchi09,Devecchi10,Devecchi12,AN14,Katz15, Sakurai17}
or throughout gas-driven core-collapse \citep[][]{Davies11,Lupi14},
and by massive BH seeds with $[10^4-10^6]~M_\odot$, 
the so-called direct collapse BHs (DCBHs). DCBHs are supposed to form
via the rapid collapse of metal poor gas clouds in which star formation is prevented 
either by high velocity galaxy collisions, or by intense fluxes of 
$\rm H_2$ photo-dissociating photons \citep[e.g.][]{BL03, LodatoNatarajan06, 
SpaansSilk06, Volonteri08, IO12, Latif13a, Latif13b, Latif14b, Visbal14c,
Sugimura14, Agarwal14, Inayoshi15, Regan17}.

In a previous work \citep[][hereafter V16]{V16} we explored the relative role of
Pop~III remnant BHs and DCBHs in the formation pathways to the first SMBHs considering 
an Eddington-limited BH accretion scenario. We used a semi-analytic, data-constrained, cosmological 
model called \gamete \, aimed at study the formation of high redshift quasars 
and their host galaxies \citep[][]{V11,V14,V16}. 
In V16 we also showed that the condition for the formation of DCBHs are very tight, so that they form
in a small number of halos (from 3 to 30 depending on the merger history) and in a very narrow 
redshift range $(16<z<18)$. 
Still, the Eddington-limited growth of a SMBH of $>10^9 \, M_\odot$, as inferred for SDSS J1148+5251, 
relies on the formation of these few massive seeds.

The aim of this work is to investigate to what extent the properties of BH progenitors and their 
host galaxies can be used to disentangle the nature of the first BH seeds, before their memory is lost
along their cosmological evolution.
Here we extend the analysis presented in V16, exploring additional physical properties of the BHs and 
their hosts. In a companion paper (Valiante et al. 2017c), we will make observational predictions for
the most promising candidates.
Our approach has the advantage of following the evolution of light and heavy seed progenitors along 
the same hierarchical history (i.e. the two BH seed formation channels are not mutually exclusive), 
starting from the epoch at which the first stars form ($z\sim 24$, in our model) 
down to $z\sim 6$, when the observed quasar is eventually assembled.
Moreover, the properties of the BH seeds population are self-consistently related to the evolution of their host 
galaxies: mass, number, redshift distribution and growth history are regulated by the build up
of the UV radiation field, by the metals and dust pollution of the ISM and IGM and by the effect 
of stellar and AGN-driven winds.

The paper is organized as follows. In Section 2 we briefly summarize the main features of \textsc{GAMETE/QSOdust}. 
In Section 3 we present the properties of the sample of accreting light and heavy seeds as predicted by our model. 
Finally, we discuss our results drawing the conclusions of the study in Section 4.

\section{BH formation and evolution model}\label{sec:model}
Here we briefly introduce the semi-analytic model \gamete, aimed at studying 
the formation and evolution of high redshift quasars and their host galaxies
at $z>5$.
We refer the reader to \citeauthor{V11} (\citeyear{V11,V14, V16}) 
for a full description of the code.

The model succesfully reproduces the BH mass and the properties 
of the quasars host galaxies, such as the mass of gas, metals and dust and the star formation 
rate.  In this analysis we select the quasar SDSS J1148+5251, observed at $z=6.4$, as our 
target. This is one of the best studied objects at $z>6$ and it is powered by a SMBH of $(2-6)\times 10^9$ \msun \, 
(\citealt{Barth03, Willott03}). We summarized the other main observed properties of this quasar 
and its host galaxy in \citet{V11} and \citet{V14}. 

With the same model we have investigated the nuclear BHs-host galaxies co-evolution histories for a sample of 
$z>5$ quasars, reproducing their observed properties \citep[][]{V14}.

\subsection{Hierarchical merger histories}\label{subsec:MTs}

We developed a Monte Carlo algorithm, based on the Extended Press-Schechter theory (EPS),
to reconstruct several hierarchical merger histories of the $10^{13} \, M_\odot$ 
host dark matter (DM) halos.
{
As introduced in V16, we also extended the DM halos mass spectrum to 
mini-halos\footnote{Halos with 
virial temperature in the range $1200~{\rm K} \le T_{\rm vir}< 10^4$ K.} which are
resolved down to redshift $z \sim 14$ along the merger trees. 
These are the birth places of the first generation of stars (Pop~III stars) 
and at $z \gtrsim 17$ represent the dominant population among DM progenitors.
Their number progressively decreases at lower redshift, down to $z \sim 14$, below which
more massive halos, with $T_{\rm vir}\geq 10^4$ K (called $\rm Lyman-\alpha$ cooling halos), 
dominate the halo population.

The use of 
a Monte Carlo approach based on the EPS theory to simulate the 
hierarchical evolution of a $10^{13} \, M_\odot$ DM halo at $z \sim 6$ 
offers the following advantages:
it enables to
{\it (i)} resolve early star-forming mini-halos at $z>20$, 
{\it (ii)} run several independent hierarchical histories of this ``biased'' 
region of the Universe, and 
{\it (iii)} sample a large parameter space, exploring the impact of poorly 
constrained physical processes. 
This combination of features can not be found in current numerical simulations.
In fact, simulations can not resolve large and small scales at the same time:
small scale simulations allow to follow the physical processes accurately (for instance the combination of feedback effects in star forming mini-halos) but suffer from poor statistics, while large scale simulations provide a larger statistics but at the price of not capturing some fundamental physical processes, that are crucial to understand early BH seeding \citep[see e.g.][for a discussion]{Habouzit16hydro}.

However, the use of an analytic merger tree comes at the price of lacking the information on the halo spatial distribution.
In Section~\ref{sec:discussion}, we will discuss the consequences of this limitation
comparing our results with independent studies. 
}

\subsection{Evolution of progenitor galaxies}
The evolution of each progenitor galaxy along the merger tree, is determined by star formation, 
BH growth and stellar+AGN feedback processes.

The star formation history (SFH) of galaxies along the merger tree is described 
as a series of quiescent episodes of star formation and/or major merger-induced bursts\footnote{We define 
as major the mergers among halos with DM mass ratio $\mu>1/4$, where $\mu$ is 
the ratio of the less massive over the most massive halo of the system.}.

The nuclear BHs can grow via both mergers with other BHs and accretion
of gas in an Eddington-limited regime (i.e. the gas accretion rate does not exceed
the Eddington value). In major mergers pre-existing BHs are assumed to coalesce, 
following their host galaxies. 
Conversely, in minor mergers only one
of the two BH is assumed to settle in the center of the galaxy. The less massive 
BH of the pair is left as a satellite (see e.g. \citealt{TH09}) and we do not follow
its evolution. Gas accretion onto the central BH is described by the 
Bondi-Hoyle-Lyttleton (BHL) rate with an additional (multiplicative) free parameter, 
$\alpha_{\rm BH}$ usually adopted in both semi-analytic models and numerical simulations
to account for the increased density in the inner regions around the BH \citep{DiMatteo05}\footnote{In \textsc{GAMETE/QSOdust} we assume that the gas mass of the galaxy is 
distributed within the virial radius of the DM halo following an isothermal sphere 
profile with a flat core. Under this assumption, the gas density at the Bondi radius is 
underestimated. In addition, the BHL accretion rate strongly depends on the sound 
speed of the local material, but we do not track the temperature of the cold gas, 
nor that of the accreting material around the BH. The BH accretion efficiency 
$\alpha$ enables us to overcome these limitations \citep[see e.g.][for a discussion]{BoothSchaye09}.}.
The different values adopted for this parameter are given in \citet{V11} and \citet{V14}.
In \citet{V16} a value $\alpha_{\rm BH} = 50 $ enables us to match the observed BH mass 
and estimated accretion rate of SDSS J1148+5251.

The progressive pollution of the interstellar medium (ISM) with metals 
and dust, produced by Asymptotic Giant Branch (AGB) stars and Supernovae 
(SNe), is computed consistently with the stellar evolutionary time scales
\citep[lifetimes; see][for details.]{V14}. In addition we follow dust enrichment 
in a two-phase ISM \citep{deBen14, V14}: dust grains residing in the hot, diffuse 
gas can be destroyed by expanding SN shocks, while in the cold, dense clouds dust 
can grow in mass by accretion of the available gas-phase elements onto the 
grains surface. 

{
The UV radiation from both stars and AGN is computed according to the emissivity 
properties of each source. 
For Pop~III stars we adopt the mass-dependent emissivities presented by 
\citet{Schaerer02} and compute the age-metallicity-dependent emissivities for Pop~II 
stars using the model by \citet{BC03}. 
Following \citet{Volonterignedin2009}, the emissivity of accreting BHs is inferred by modelling 
their SEDs with a multicolor disc spectrum plus a non-thermal power-law component 
\citep[see][for details]{V16}.

We compute the LW blackground within a comoving volume of $50$ Mpc$^3$ (our biased, 
high-density region at $z > 6$), following \citet[][]{HM96} in the dark screen 
approximation and accounting for intergalactic absorption using the modulation 
factor given by \citet[][]{Ahn09}.
As discussed in \citet[][]{V17a}, despite the lack of spatial information, in the highly 
biased volume that we simulate, the flux level that we predict for the LW background 
is comparable to the maximum local flux found by \citet{Agarwal12}. We will further
discuss this point in Section~\ref{sec:discussion}.
}

The effect of stellar and BH mechanical feedback is included in the form of 
energy-driven winds which are able to trigger galaxy-scale gas outflows 
(including metals and dust) polluting the intergalactic medium (IGM).
We assume that a fraction $\epsilon_{\rm w,SN}=2\times 10^{-3}$
and $\epsilon_{\rm w,AGN}=2.5\times 10^{-3}$ of the energy 
released by SN explosions and BH accretion process, respectively, is responsible 
for launching gas outflows (see V16 for details).\\
Although they have similar wind efficiencies, in \citet{V12} we show that the 
SN-driven winds have a minor effect on the BH-host galaxy co-evolution history 
which is instead regulated by the AGN feedback, as observed in 
SDSS J1148+5251 \citep{Maiolino12,Cicone15}.

\subsection{Forming the first stars and BHs}

As discussed in V16, the combined levels of chemical enrichment (in situ- and/or 
infall-driven) and of the UV flux to which the host galaxy is exposed, 
regulate star formation and set the conditions for DCBHs formation. 

{ 
In \textsc{GAMETE/QSOdust}, we describe both in-situ and external (or ex-situ) pollution. 
The first dominates at high redshift and the second progressively becomes important 
and ultimately leads to a termination of Pop~III star formation at $z\sim 16$ (on average,
see Section~\ref{sec:discussion} for a discussion).  
}

Whether stars can form in mini-halos and what is the efficiency of star formation is 
dictated by the halo properties (virial temperature, redshift and ISM metallicity) and by
the photo-dissociating radiation illuminating the halo (see Appendix A in V16 for details).

Radiative feedback also determines the minimum mass of star-forming halos.
This threshold mass increases with redshift, first as a consequence of $\rm H_2$ 
photo-dissociation 
\citep[e.g.][and Appendix A in V16]{Omukai01, Mach01},
then, at lower redshifts, due to ionization which can lead to photo-evaporation
of less massive halos (e.g. \citealt{BL99}, \citealt{Shapiro04}, \citealt{SM13}).

In addition, chemical feedback determines the type of stellar population that can form in a galaxy.
We assume that Pop~III stars in the mass range $[10-300] \, M_\odot$ form in metal poor halos, 
as long as the ISM metallicity is lower than the critical threshold $Z_{\rm cr}\sim 10^{-4} \, Z_\odot$ 
\citep[][]{Schneider02, Schneider03, Schneider12}. Conversely, 
second generation stars (Pop~II), with a mass spectrum shifted towards less massive stars,
$[0.1-100] \, M_\odot$, form out of gas enriched above this critical value.
The stellar initial mass function (IMF) is assumed to have the Larson functional
form \citep{Larson98}: $\Phi(m)\propto m^{\alpha-1} e^{-m_{ch}/m}$, where $\alpha=-1.35$
and $m_{ch}=20 \, (0.35)$ is the characteristic mass for the Pop~III top-heavy (Pop~II standard)
IMF\citep[][]{deBen14}.   

In low-efficiency star forming halos, we stochastically sample the Pop~III IMF. 
Following each star formation episode, we randomly select stars in the $[10-300] \, M_\odot$ mass 
range until we reach the total stellar mass formed in the burst.
As a result only few 
less massive Pop~III stars can form in mini-halos, while the intrinsic 
top-heavy IMF is fully sampled if a total stellar mass larger than $10^6\,  M_\odot$ is produced.
This approach enables us to avoid overestimating the contribution of Pop III stars to radiation 
emission and chemical enrichment.

\subsubsection{Seeding prescription}
Following this stochastic procedure to sample the Pop III IMF, we form light seeds when we extract 
massive stars in the range  $[40-140] \, M_\odot$ and $[260-300] \, M_\odot$. 
We assume that only 
the most massive among BH remnants formed in each burst settles in the galaxy center where it can 
accrete gas and grow. 

We compute the time evolution of the cumulative LW emission, $J_{\rm LW}$, coming from all the 
emitting sources (the progenitor galaxies and BHs of SDSS J1148+5251) along the merger trees 
(see section 2.2 in V16). 
We then assume that a DCBH can form when $J_{\rm LW}>J_{\rm cr}$, where  
$J_{\rm cr}=300 \times 10^{-21} \, \rm erg/s/Hz/cm^2/sr$, in our reference model\footnote{The value 
of this critical threshold ranges from few tens up to $10^3-10^4$ times 
$10^{-21} \, \rm erg/s/Hz/cm^2/sr$ depending on the model or simulation adopted\citep[see e.g.][]{Shang10,
WolcottGreen2011,Latif14a,Latif14b,Regan14,Hartwig15}. We refer the reader to 
the recent review by \citet{V17a} for a detailed discussion.}.
This critical level is exceeded at $z \sim 16-18$ (depending on the particular merger tree) so that
a massive seed of $10^5 \, M_\odot$ is assumed to form in poorly enriched ($Z_{\rm ISM}<Z{\rm cr}$)
$\rm Ly\alpha$ cooling halos.

In-situ or external metal pollution determines the Pop~III to Pop~II transition epoch, $z \sim 15$. 
Below this redshift, both light and heavy seeds no longer form.

\begin{figure}
\vspace{-0.5cm}
\includegraphics [width=7.5cm]{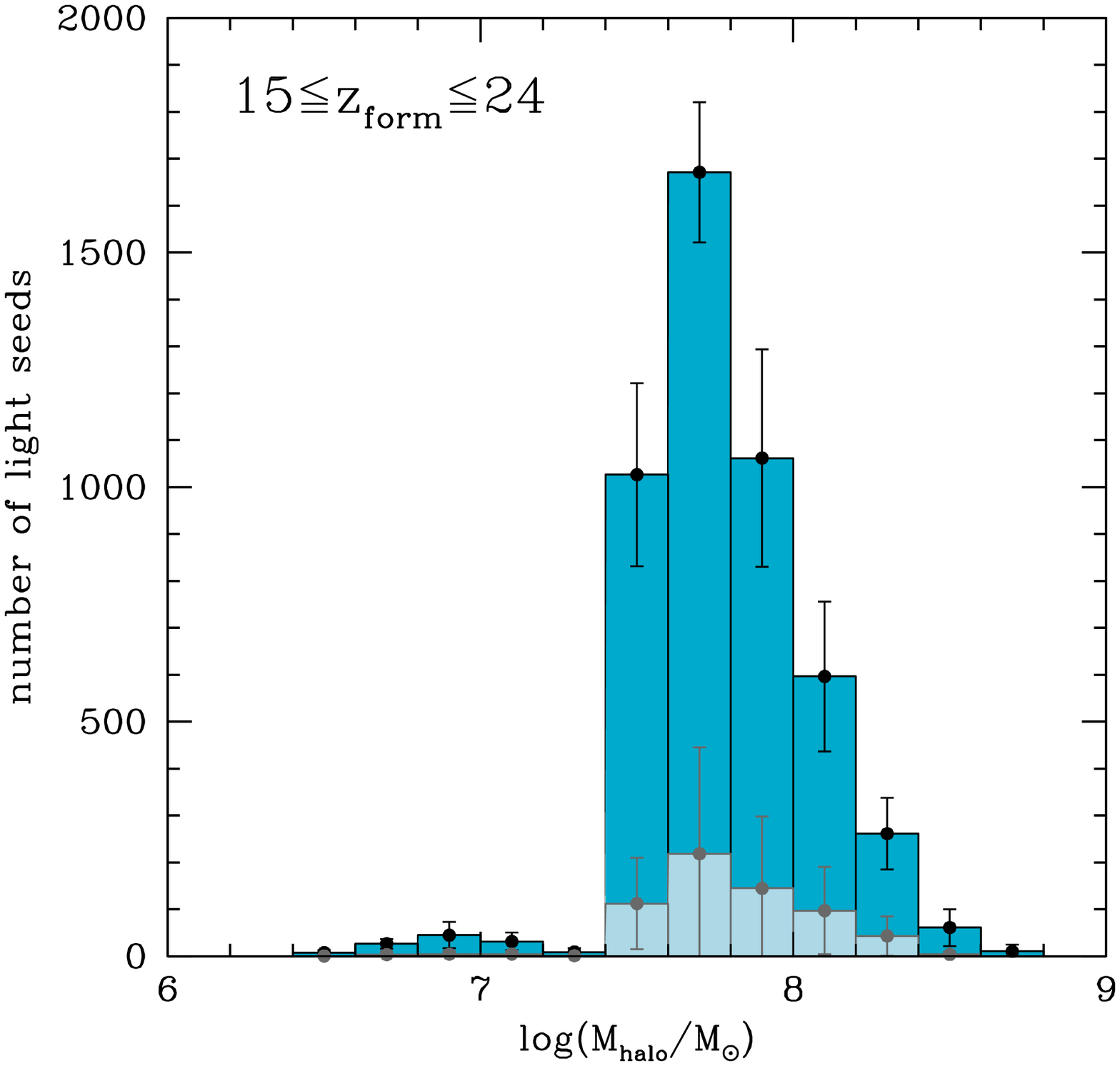}
\includegraphics [width=7.5cm]{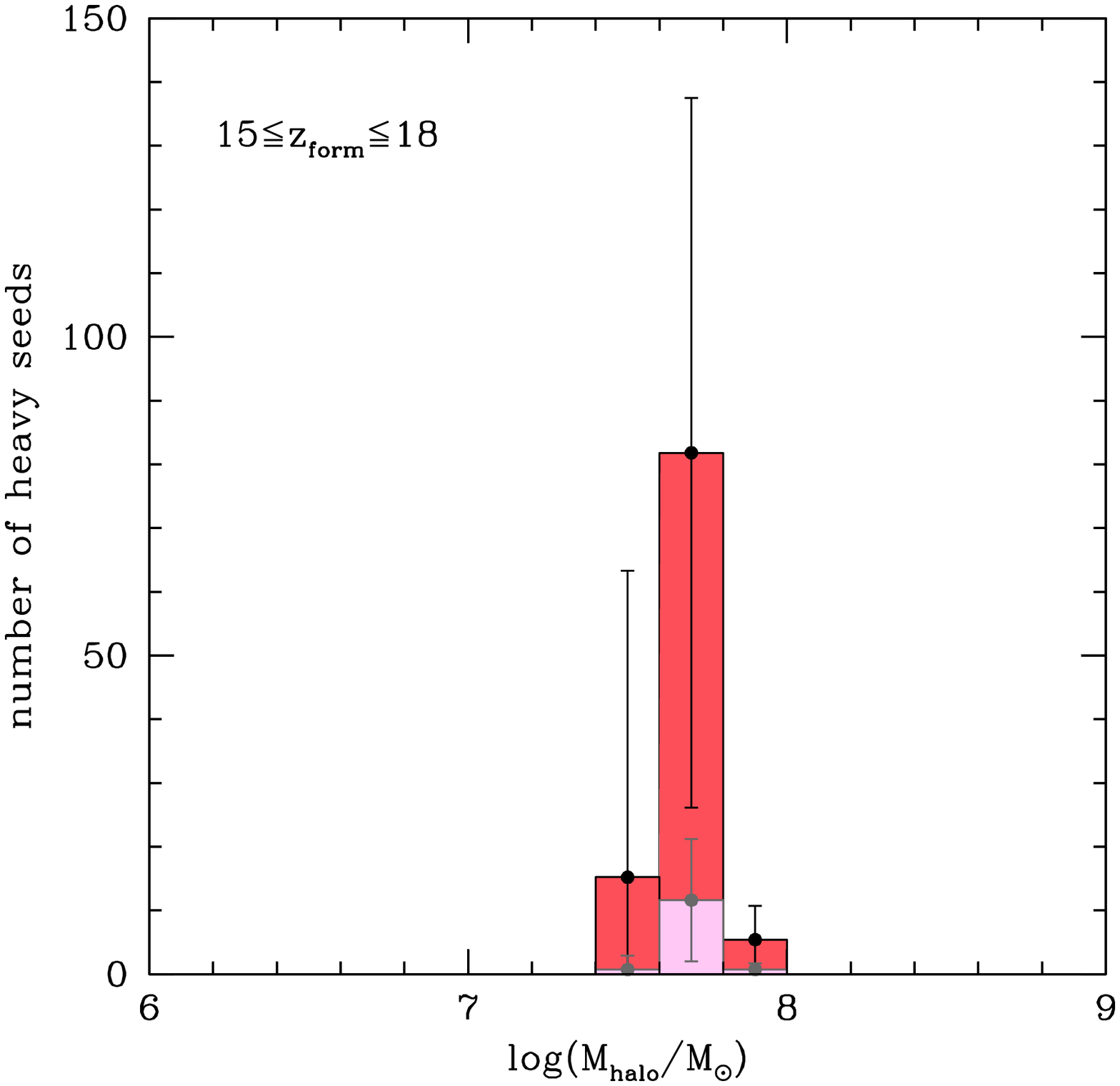}
\caption{Mass distribution of DM halos hosting light (upper panel) and heavy (lower panel) 
BH seeds. The total number of BH seeds is shown with darker colors. Ligther histograms show the 
subsample of seeds that directly classify as SMBH progenitors (see V16 for details). 
The average range of formation redshifts is also reported in each panel.} 
\label{fig:haloMass} 
\end{figure} 

\begin{figure*}
\centering
\includegraphics [width=7.5cm]{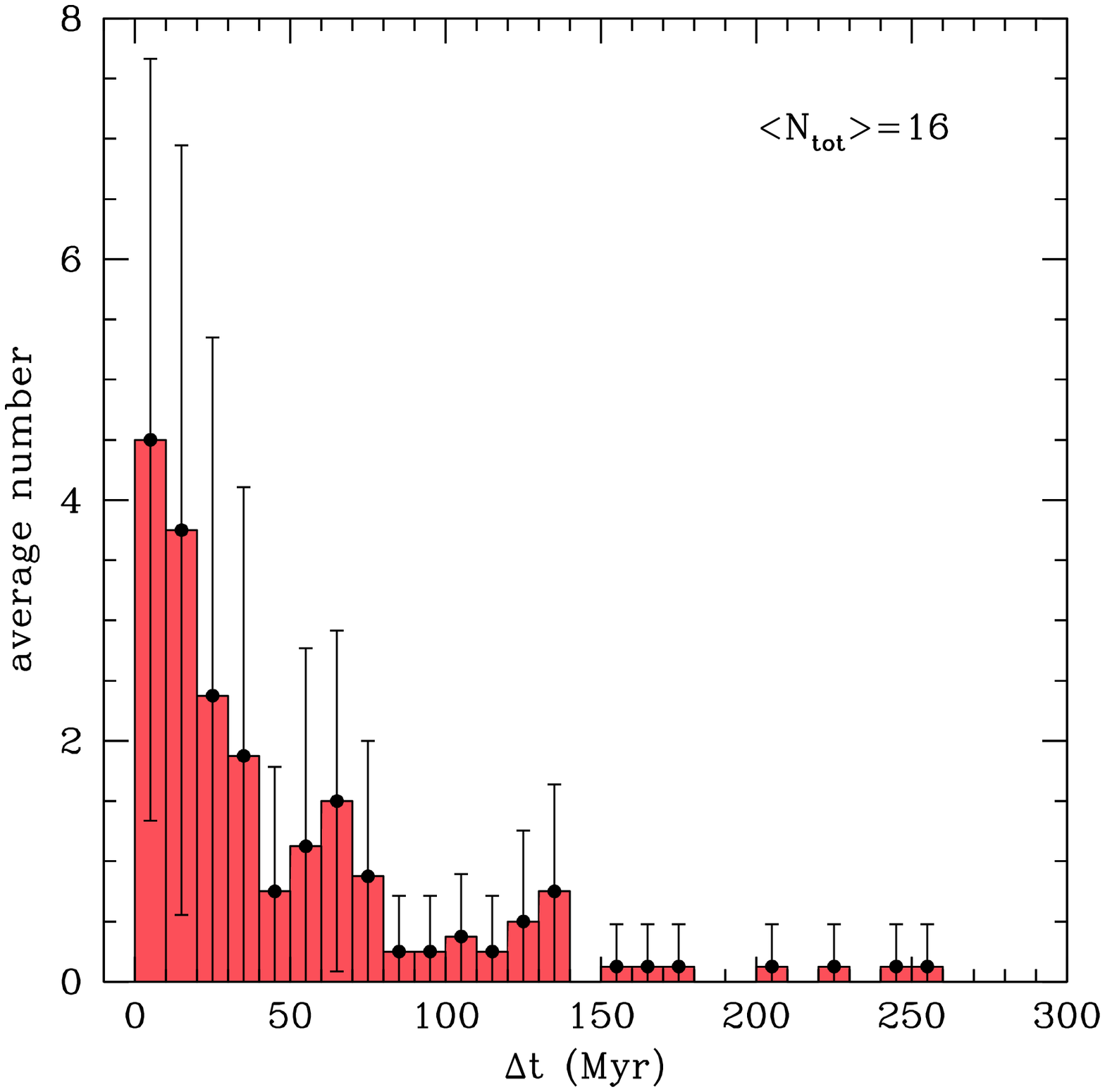}
\includegraphics [width=7.5cm]{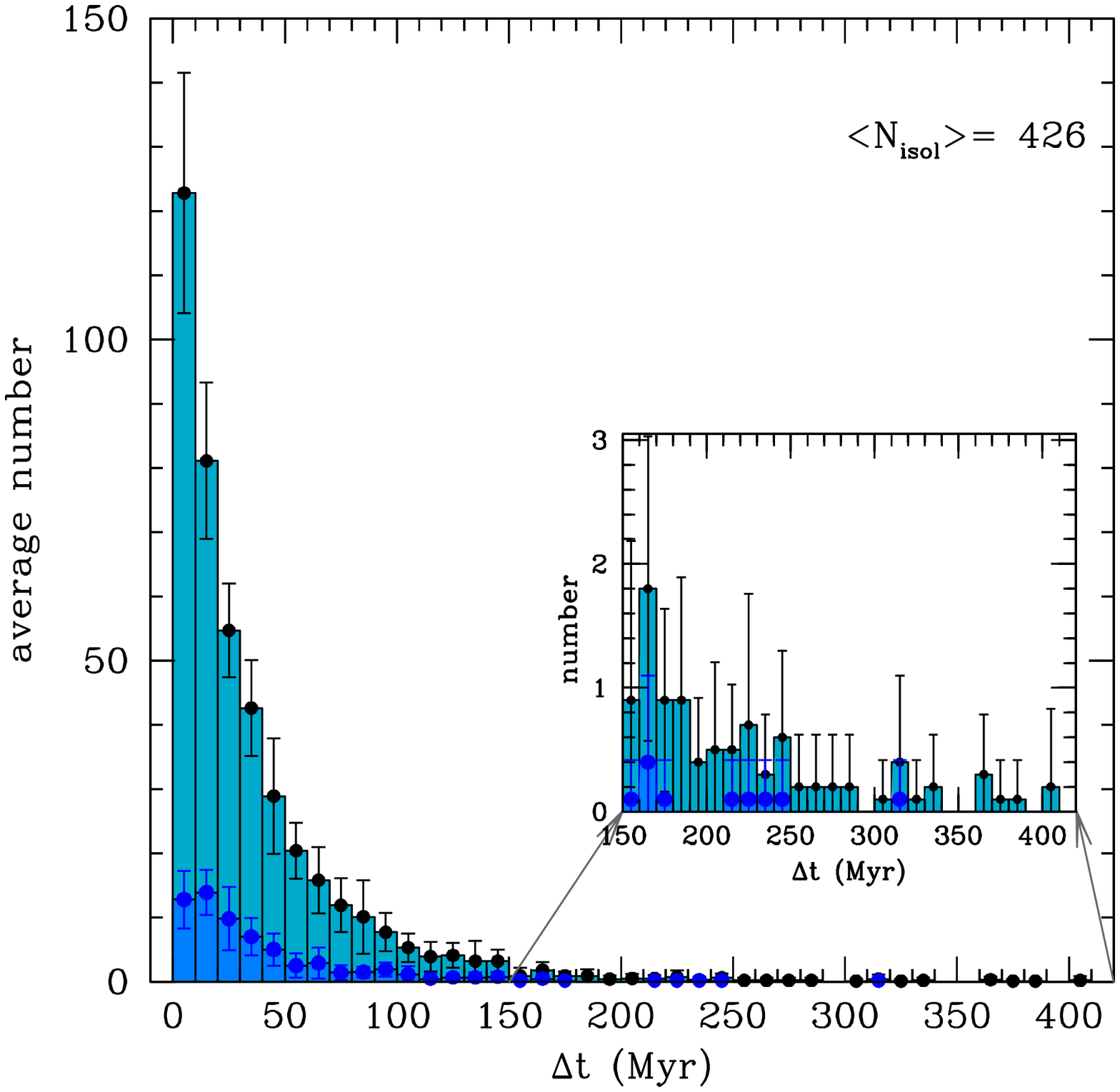}
\caption{Distribution of isolated heavy (left) and light (right) seeds as a function of  (rest-frame) time.
The labels indicate the average number of IHS and ILS formed in 10 different merger trees.
In the right panel, the superimposed bluer histogram shows the number of ILS with 
masses $>10^3 \, M_\odot$ and the small inserted plot enlightens the long-living seed tail,
at $\rm \Delta t>150\,Myr$.} 
\label{fig:isolatedseeds} 
\end{figure*} 

\section{RESULTS} 
Here we investigate the 
statistical properties of BH seeds and their host galaxies.
In V16 we discussed the evolution and properties of BH progenitors of a $z\sim 6$ SMBH,
i.e. excluding satellite BHs emerging from minor galaxy merger events.
The analysis presented here considers the entire high$-z$ BH population that we find in the simulations, 
independently of whether they will contribute to the formation of the SMBH mass at $z\sim 6$.

In what follows histograms and data points will show quantities obtained by 
averaging over 10 different merger tree realizations, with $1\sigma$ errorbars, 
unless otherwise stated. 

\subsection{Properties of BH seeds and their hosts} 
\label{sec:sample}
Fig.~\ref{fig:haloMass} shows the mass distribution of DM halos hosting BH seeds formed along 
the merger history of the halo hosting SDSS J1148+5251. 
On average, light and heavy seeds are predicted to form in halos with 
comparable mass and at comparable redshifts. The average number of light BH seeds 
forming in mini-halos with $[10^6 - 10^7] \, M_\odot$ (at $z>20$, see V16), is small. 
This is a consequence of the LW feedback which strongly limits the efficiency of star formation
in these halos. Hence, the bulk of light seeds form at redshift $15 < z_{\rm form} < 20$
in metal poor ($Z \le Z_{\rm cr}$) Ly$\alpha$-cooling halos.
On the other hand, the combined effect of chemical and radiative feedback 
enables the formation of heavy seeds (bottom panel) only in a relatively small number of 
halos, with masses in the range $[5\times 10^7 - 10^8] \, M_\odot$ and at redshifts $15<z_{\rm form}<18$.

\subsection{BH seed hosts evolving in ``isolation''}
\label{sec:BHseedsiso}

In what follows, we analyze the properties of BHs as long as their host galaxies evolve in 
``isolation'', namely from the epoch of the BH seed formation down to the redshift of the 
first merger with a companion galaxy. 
During this phase, the galaxies only interact with the external medium, via gas
infalls and outflows, without experiencing any major or minor merger.
We will refer this class of objects as \textit{isolated seeds} hosts.
The aim of this analysis is to bring out the distinguishing features of light and 
heavy seeds, which may help us discriminating their observational signatures.

Fig.~\ref{fig:isolatedseeds} shows the distribution of 
isolated light and heavy seeds (hereafter ILS and IHS, respectively) as a function of (rest-frame) time. 
Labels in the two figures indicate the total number of galaxies hosting IHS and ILS, averaged over 10
different merger trees\footnote{In an isolated halo a heavy BH seed grows only via gas 
accretion while a light seed can also merge with other Pop~III BHs that may form in subsequent 
starbursts occurring in the same halo (see V16 for details). 
For the purpose of the present analysis, we consider as the effective light seed only the first 
Pop~III BH formed in the halo, when counting the number of ILS and defining the duration of the  
isolated phase, $\Delta t$.}. These represent $80~(98)\%$ of light (heavy) seed hosts at $z > 10$.

The majority of IHS (ILS) hosts merge with another galaxy within the first 
$\sim 50$~Myr, and only a small fraction of systems, $14~(7)\%$, remains isolated for more 
than 150~Myr, as indicated by the tail of the distributions extending to $\Delta t \sim 250~(400)$~Myr. 

\begin{figure}
\centering
\includegraphics [width=8cm]{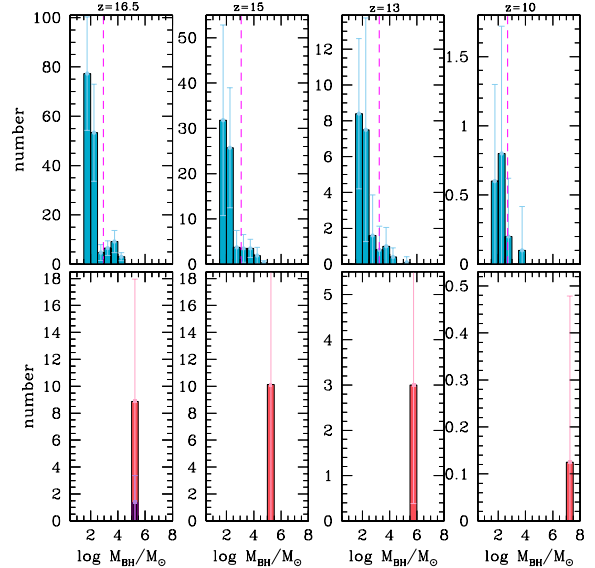}
\caption{Mass distribution of accreting light (upper panels) and heavy (lower panels) BH 
seeds at 4 different redshifts: z=16.5, 15, 13 and 10. 
The dashed lines in the upper panels indicate the average mass of ILS at the corresponding $z$.
} 
\label{fig:massdistriBHseeds} 
\end{figure}

During their evolution, ILS and IHS mantain different BH masses and ILS are
less massive than IHS.
This is shown in Fig.~\ref{fig:massdistriBHseeds} where the histograms represent the mass 
distribution of ILS and IHS at $z= 16.5, 15, 13,$ and 10.
The redshift range has been chosen in order to encompass a period of time during which the evolution of 
both IHS and ILS can be compared. At $z=16.5$ we find newly planted DCBHs 
together with a number of evolving DCBHs formed at earlier times, so that we can capture 
the host galaxies properties in both phases.
Below $z=10$, it is increasingly difficult to consider light and heavy seeds as two independent 
populations, due to mergers along their cosmological evolution (see Section~\ref{sec:BHseedmerged}).

The difference in the mass spectrum of the two populations grows with time. 
In fact, at $z < 15$ when the mass growth is mostly driven by gas accretion, the BHs grow
$\propto M_{\rm BH}^2$ when gas accretion
occurs at the Bondi rate, and $\propto M_{\rm BH}$ when it occurs at the Eddington rate. 
Hence, by $z \sim 10$ the mass of 
IHS is $[10^7 - 10^8] \, M_\odot$, i.e. about two orders of magnitude larger than the upper mass 
limit of ILS at the same redshift.
The statistics also include a small fraction ($\sim 1\%$) of ILS which grow almost as fast as 
their most massive counterparts, reaching a mass $> 10^4 \, M_\odot$, up to 
$\sim 2\times 10^5 \, M_\odot$ at redshift 12-13 (see the last 2 mass bins in the third upper panel, 
from the left). 
Although they are not representative of the average ILS population, these BHs are interesting 
objects for our study, as their emissivity can directly compete with that of IHS (see Valiante et al. 
2017c).

\begin{figure*}
\includegraphics [width=7.5cm]{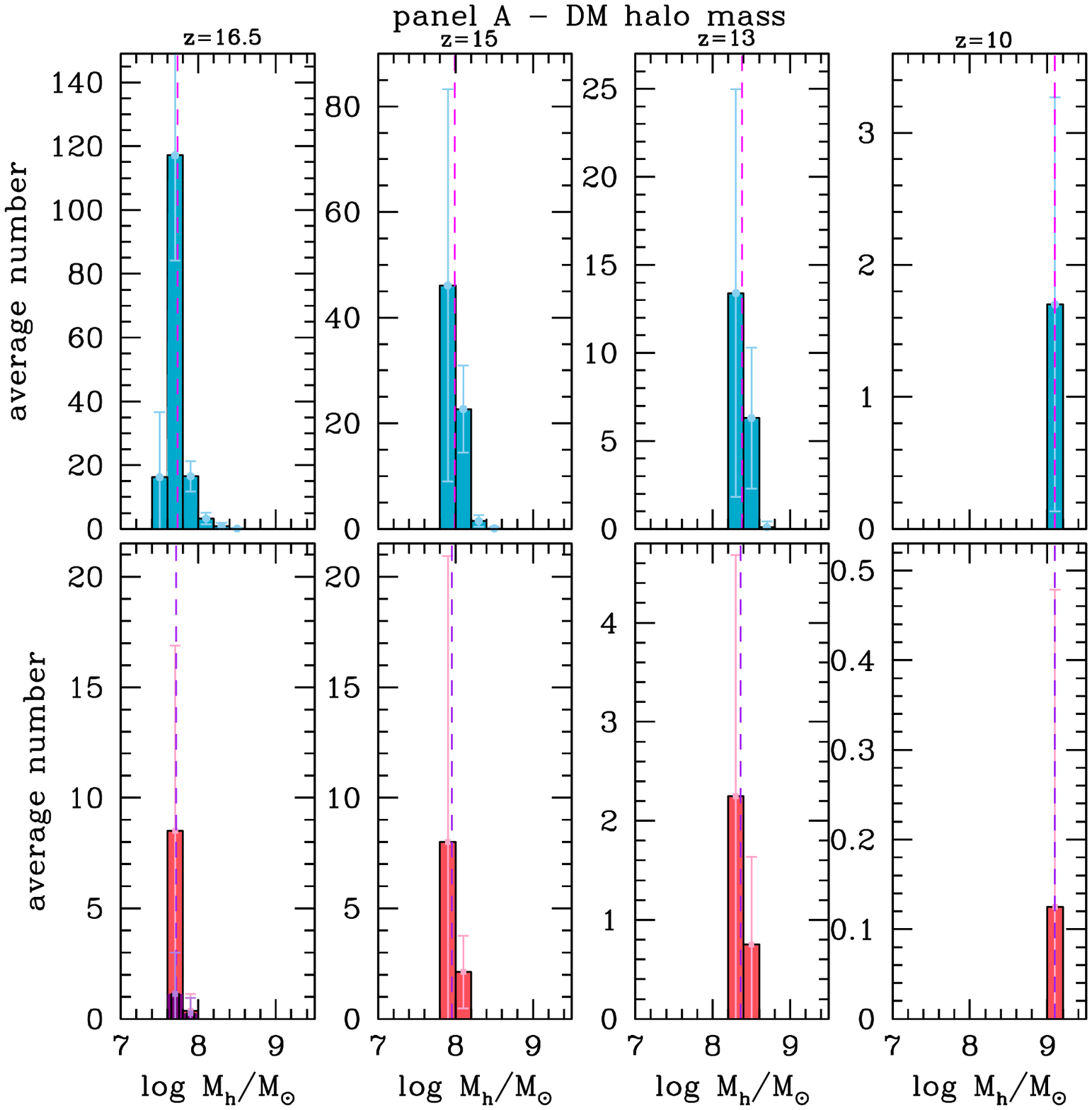}
\includegraphics [width=7.5cm]{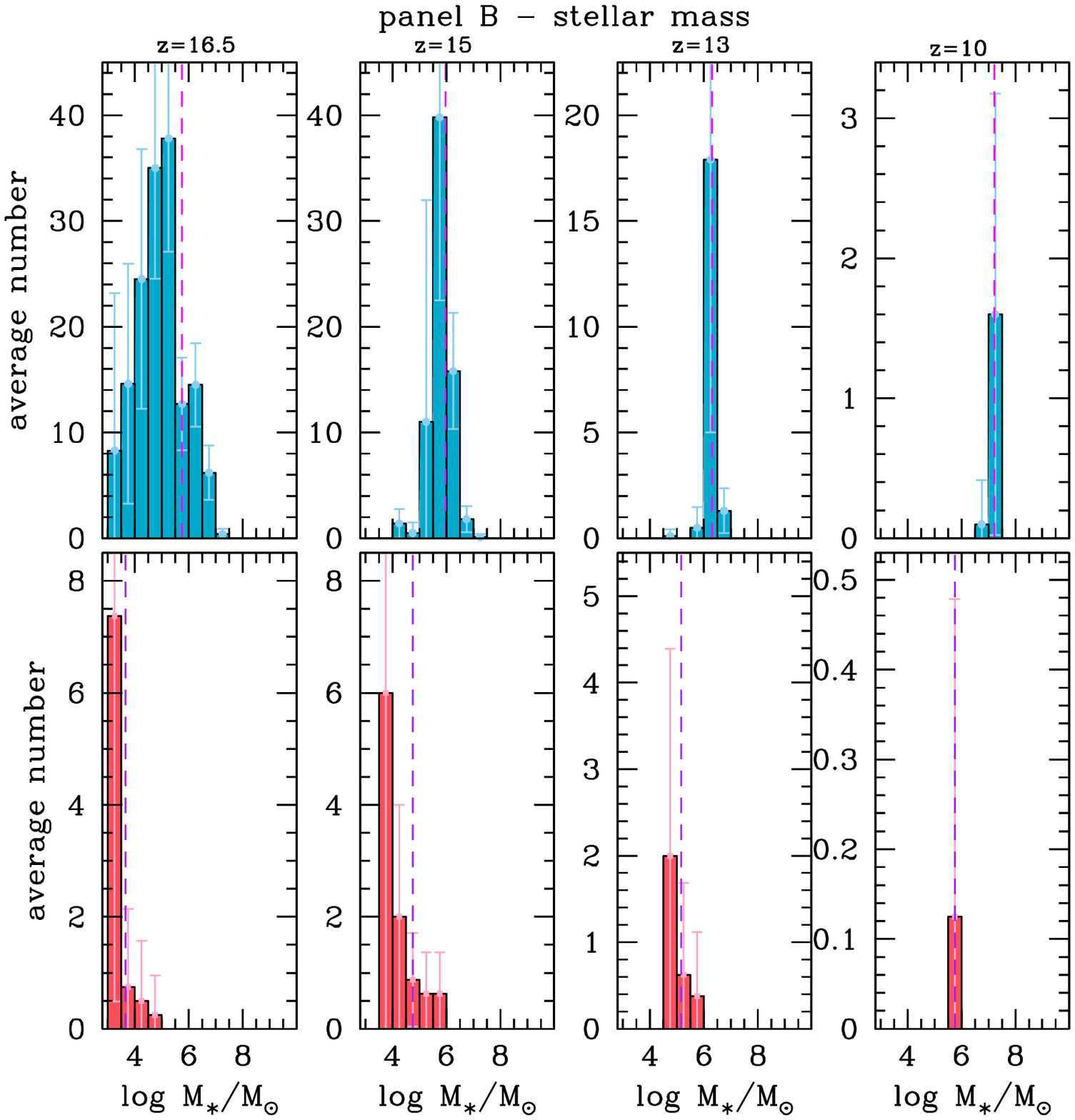}
\includegraphics [width=7.5cm]{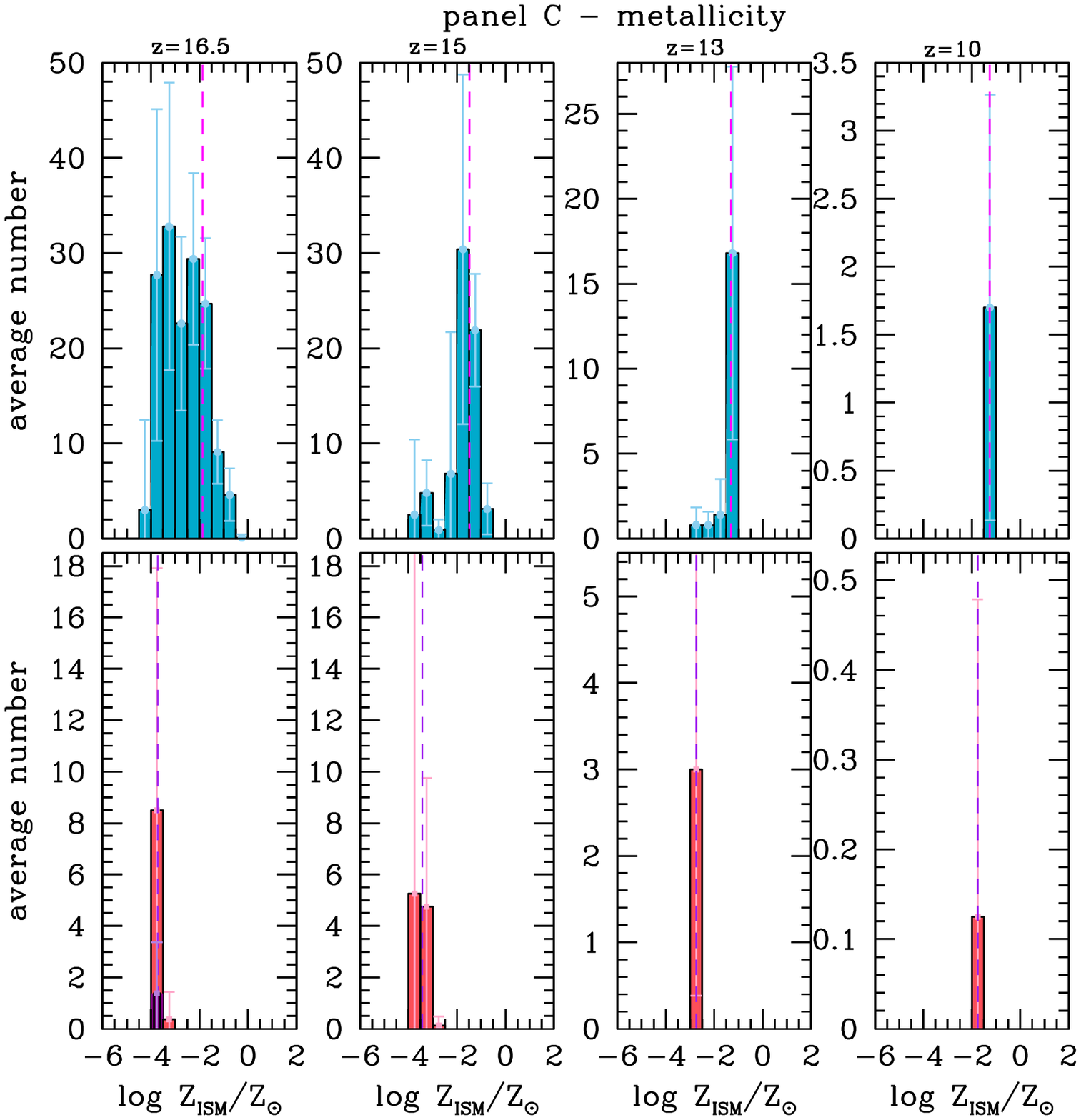}
\includegraphics [width=7.5cm]{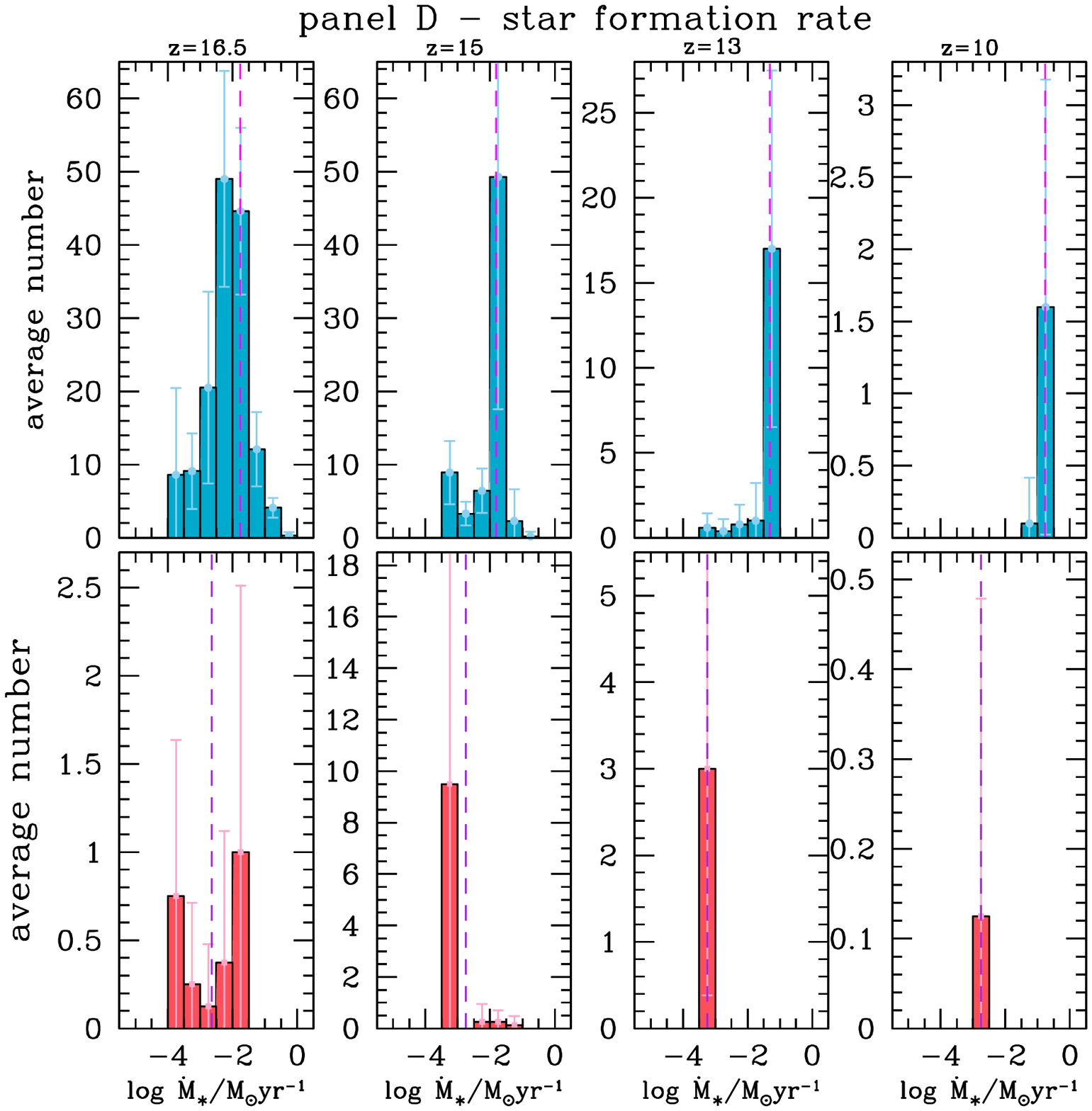}
\includegraphics [width=7.5cm]{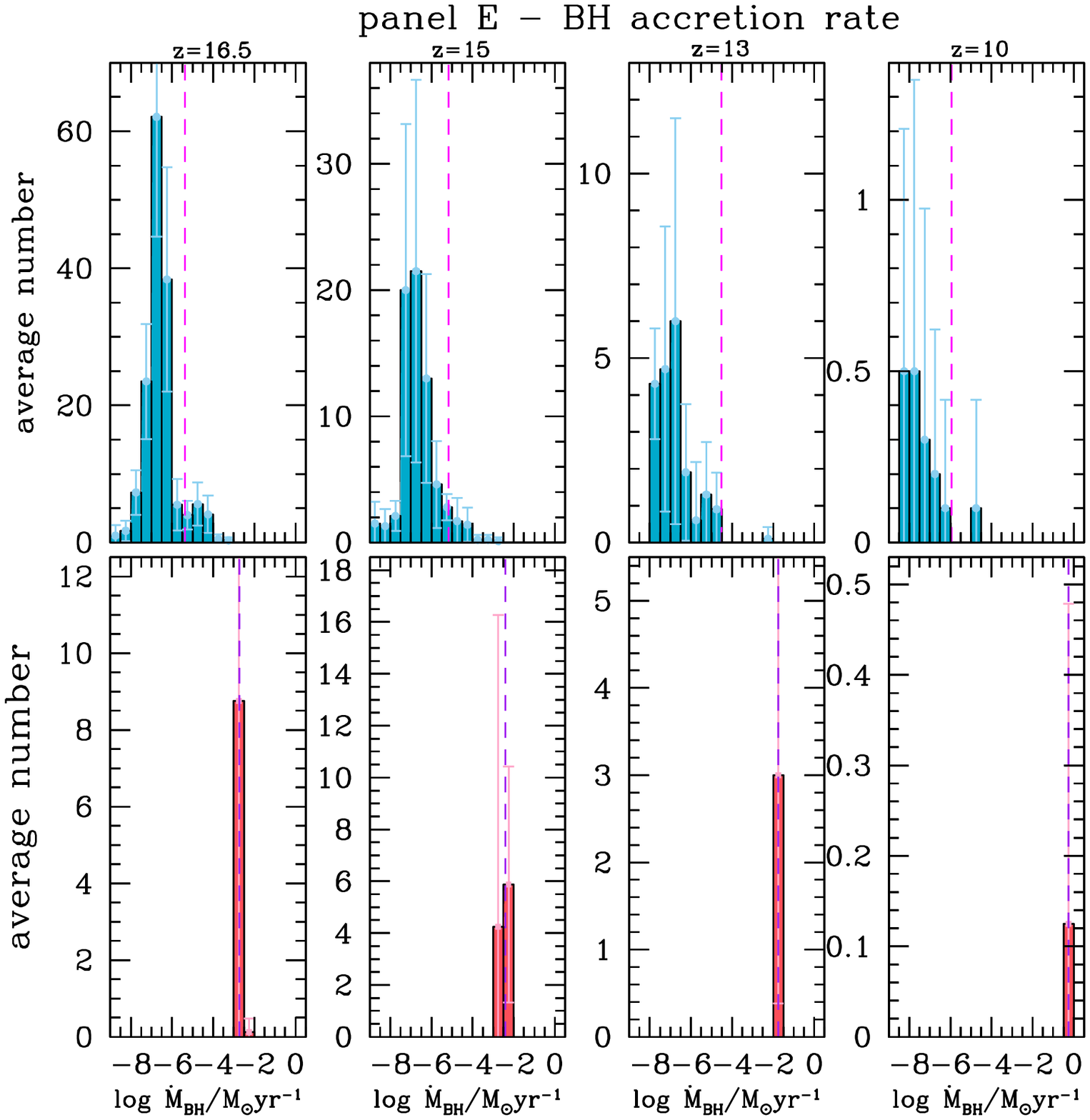}
\includegraphics [width=7.5cm]{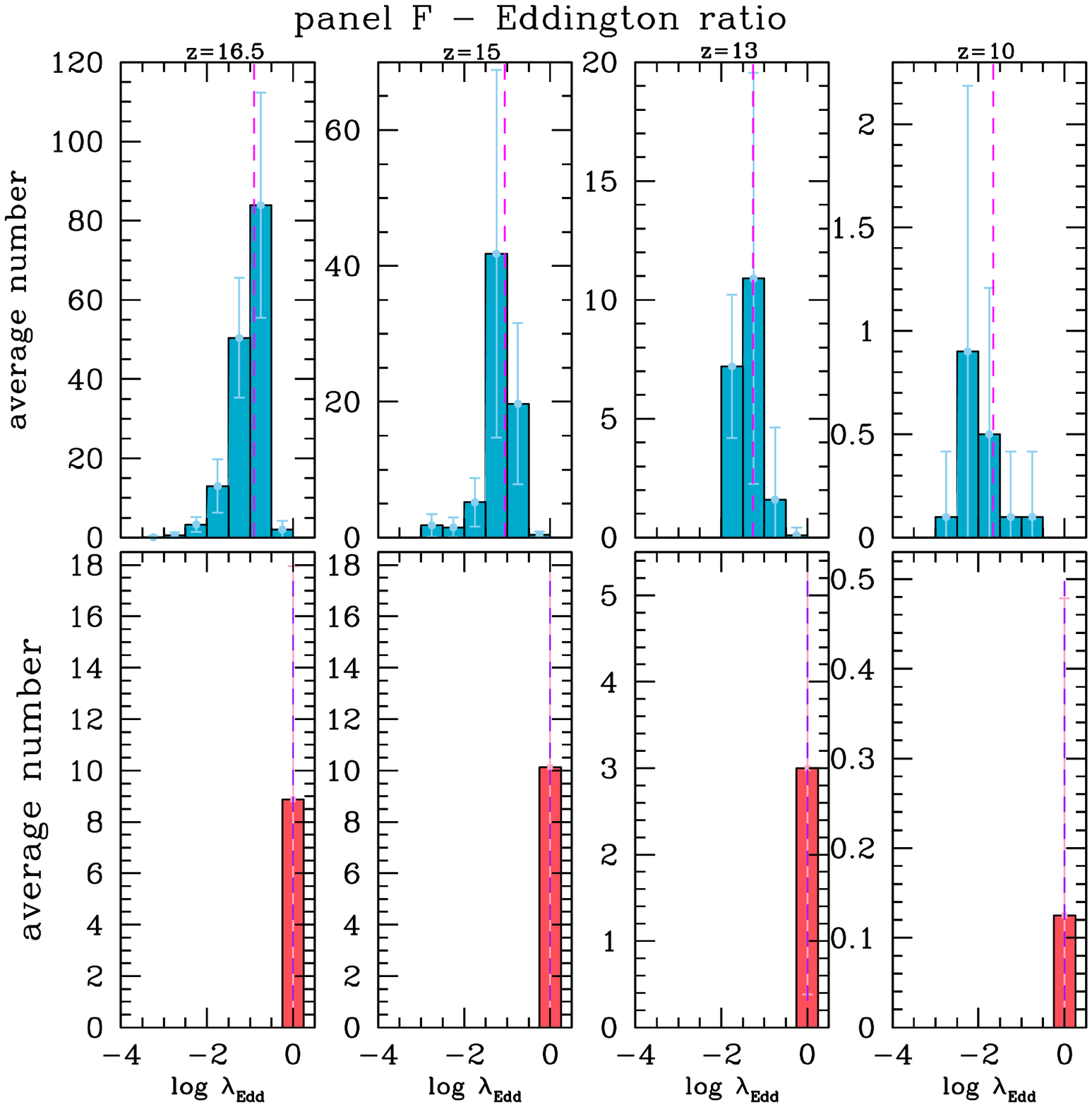}
\caption{The distribution of DM halo mass (panel A), stellar mass (panel B), 
ISM metallicity (panel C), SFR (panel D), BH accretion rate (panel E) 
and Eddington ratio $\lambda_{\rm Edd}$ (panel F) at $z=16.5, 15, 13 \rm and 10$. 
Upper, blue histograms show ILS, while lower,
red, histograms show IHS. 
Vertical dashed lines in all panels show average values.} 
\label{fig:isolatedSeedsProp} 
\end{figure*}

In Fig.~\ref{fig:isolatedSeedsProp} we show the distribution of DM halo mass (panel A), 
stellar mass (panel B), ISM metallicity (panel C), star formation rate (panel D), BH accretion rate 
(panel E) and Eddington ratio $\lambda_{\rm Edd}$ (panel F) at the same redshifts as in 
Fig.~\ref{fig:massdistriBHseeds}. Vertical dashed lines in all panels show average values.
We discuss the properties of ILS (upper panels) and IHS (lower panels) below.\\

\noindent
{\bf DM halo mass distribution}\\
As it can be seen in panel A, at $z=16.5$ both ILS and IHS 
reside in Ly$\alpha$ cooling halos with mass $\sim 5 \times 10^7 \, M_\odot$, 
with only a small fraction ($<15\%$) of ILS being hosted in more massive ones. 
The violet histogram in the first bottom panel on the left shows the newly-planted heavy 
seeds\footnote{In our model we stop 
planting heavy seeds already at redshift $z\sim 15.5$, on average (see \citealt{V16}) as a 
consequence of the fast chemical enrichment.}.
At lower redshifts, the two populations of accreting seeds continue to reside in small DM halos, 
with very similar average masses. 
In particular, at $z=10$ all seeds are found to lie in the same 
$(1 - 5)\times 10^9 \, M_\odot$ DM halo mass bin.\\

\noindent
{\bf Stellar mass distribution}\\
In terms of stellar mass (panel B), ILS host galaxies are, 
on average, more massive than the ones hosting IHS. The average stellar masses 
differ by a factor of $\sim 5-10$ at all redshifts.
Such a delayed growth of the stellar component in IHS hosts reflects the  
environmental conditions required to form the seeds as well as the more efficient
BH feedback effect. 
In ILS host galaxies, star formation starts at redshift $z>20$, with
Pop~III stars and their remnant BHs forming down to $z\sim 15$. 
On the other hand, heavy seeds begin to form at $z\sim 18$ in
Ly$\alpha$ cooling halos that are the descendants of sterile mini-halos, where
star formation has been suppressed by radiative feedback (see \citealt{V16} for a detailed discussion). 
Following the formation of the DCBH, BH accretion and feedback
prevent efficient star formation to occur, as a large fraction of the available gas mass is either accreted onto the BH or ejected by the BH-driven wind. 
This is reflected in the higher BH accretion rate (BHAR) vs star formation rate depicted in panels D and E. \\

\noindent
{\bf ISM metallicity distribution}\\
For the same reasons discussed above, the ISM metallicity of 
IHS hosts is always lower (by about the same factor as the stellar mass) than
the metallicity found in the hosts of ILSs, as shown in panel C. \\

\noindent
{\bf Star formation rate}\\
At $z=16.5$ only $\sim 30\%$ IHS hosts are star forming (see panel D).
The remaining $\sim 70\%$ do not form stars as a consequence of the strong LW flux to which 
the halos are exposed. They are either catched at DCBH formation,
or in the subsequent stages, before metal enrichment enables gas cooling and 
Pop~II star formation, overcoming the effects of radiative feedback\footnote{It is worth noting 
that enrichment of these systems is a consequence of metal- (and dust-) rich gas accretion.}. 
At $z\leq 15$ efficient BH accretion and feedback (see also panels E and F) strongly 
limit star formation in IHS hosts, enlarging the separation between the average star formation rate
of the two populations. 
As a result, at $z=10$ the star formation rate of IHS hosts is about two orders of magnitude lower than in ILS hosts. \\

\noindent
{\bf BH accretion rate and Eddington ratio} \\
The last two lower panels of Fig.~\ref{fig:isolatedSeedsProp} show the distribution
of BH accretion rate, $\dot{M}_{\rm BH}$, and Eddington ratios, $\lambda_{\rm Edd}$.
The average BH accretion rate of heavy seeds is always one order of magnitude (or more)
larger than that of light seeds (panel E) and heavy seeds grow close to or at the Eddington rate, 
with $\lambda_{\rm Edd} \sim 1$ (panel F). \\

We conclude that at $z \geq 10$ galaxies that host heavy BH seeds have smaller stellar mass and
metallicity and form stars at a smaller rate than galaxies that hosts light BH seeds. In addition, 
heavy BH seeds grow at a higher rate.

\subsection{BH seed evolution at $z<10$}
\label{sec:BHseedmerged}
\begin{figure*}
\centering
\includegraphics [width=7.5cm]{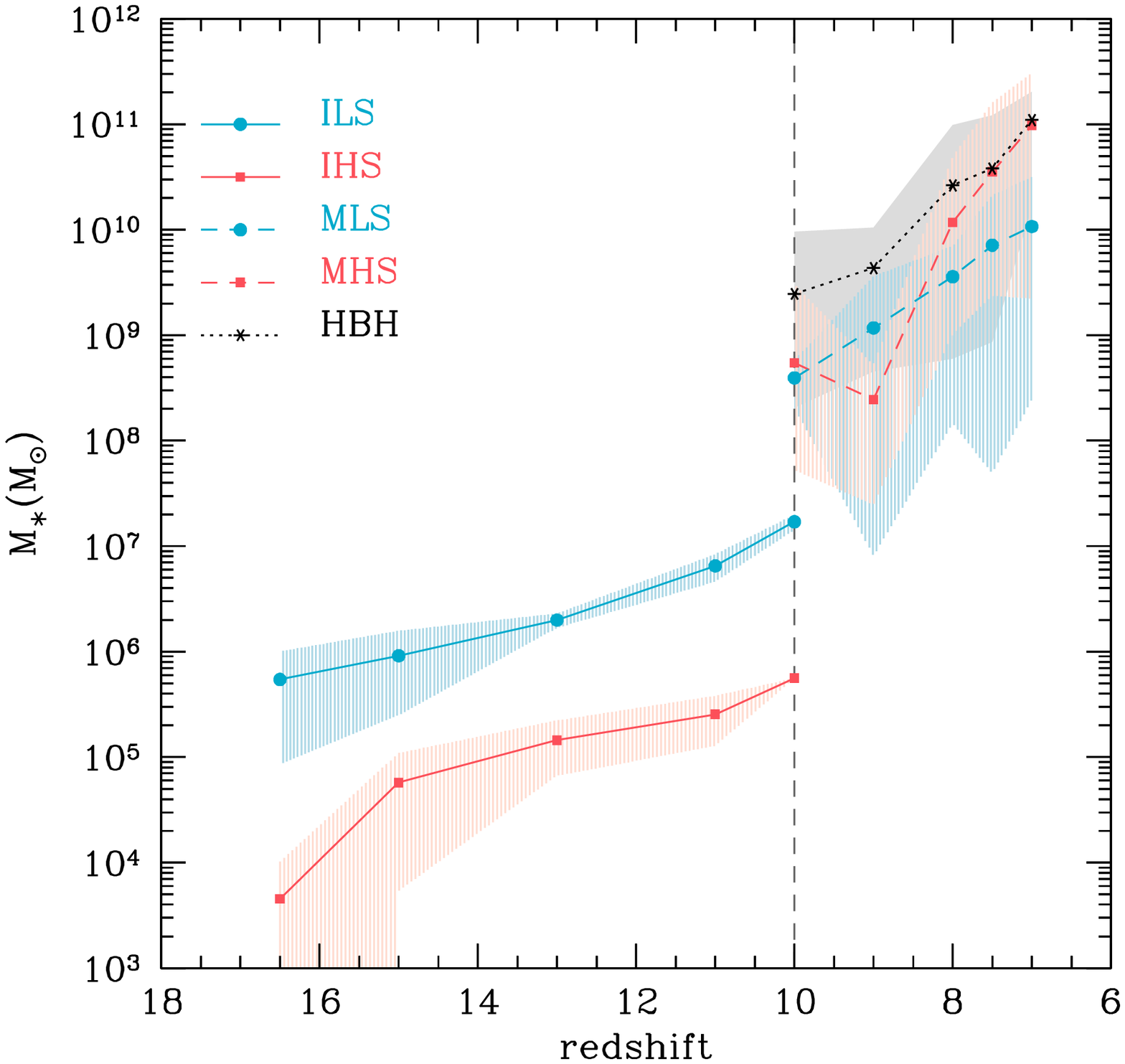}
\includegraphics [width=7.5cm]{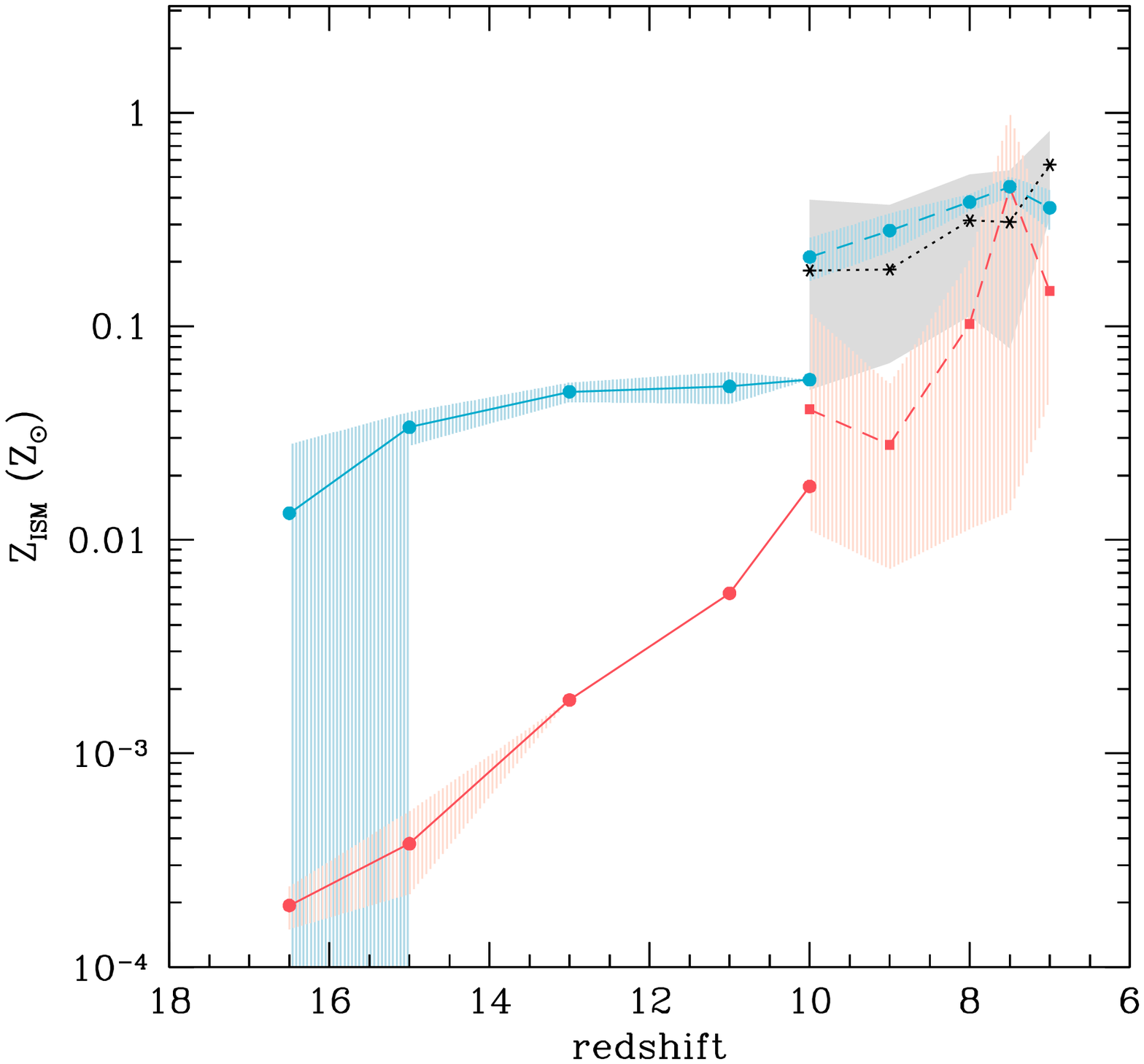}
\caption{The redshift evolution of the average stellar mass (left panel) and metallicity 
(right panel) of galaxies hosting ILS (blue circles with solid line), IHS (red squares with solid 
line), MLS (blue circles with dashed line), MHS (red squares with dashed line) and HBH (black stars 
with dotted line). The shaded regions show the $1\sigma$ dispersion, 
while the vertical dashed line indicates  $z = 10$, below which only a minor fraction of  IHS (ILS) hosts  evolve in isolation.
See text for details.} 
\label{fig:postMergerGrowth} 
\end{figure*}

At $z\leq 10$ merger events with normal, star-forming galaxies (i.e. not 
hosting nuclear BHs) or with other AGN progressively erase the differences 
in the properties discussed above, so that any trace of the BH origin is lost. 
This appears clear in Fig.~\ref{fig:postMergerGrowth}, where we show the average 
stellar mass and metallicity as a function of redshift. ILS and IHS hosts at $z \geq 10$ are shown by 
data points connected by solid lines. At $z < 10$ less than $2~(20)\%$ of IHS (ILS) hosts are 
found to evolve in isolation. 
Post-merger systems are divided into three different classes, 
according to the nature of the merging pairs:
\textit{merged heavy (light)} seeds hosts, hereafter MHS (MLS) are hosted in
halos assembled via the coalescence of an IHS (ILS) hosts with either a normal galaxy with
no nuclear BH or with another IHS (ILS) hosts;
\textit{hybrid BHs} (HBH) hosts are formed from the coalescence of IHS and ILS hosts.
The figure shows that at $z < 10$ differences between these three classes of systems 
become statistically insignificant, with the only exception of MHS metallicity, that remains
1 dex smaller than that of MLS and HBH down to $z \sim 8$.

\section{Discussion}\label{sec:discussion}

As discussed in Section~\ref{subsec:MTs}, we employed a Monte Carlo algorithm 
based on the EPS theory, with the aim of simulating the build-up of one of the 
first SMBHs at $z>6$ starting from Pop~III forming minihalos (see V16 for details). 
The main limitation of our Monte Carlo approach is the lack of spatial 
information.  
This implies that metal enrichment and UV emission from neighboring halos, 
which affect the transition from Pop~III to Pop~II stars and the number 
density of DCBHs \citep[see][and references therein]{Agarwal12, Visbal14c, 
Habouzit16hydro, Regan17}, can be modelled only in an average way. 
 
To support the robustness of our results, in this section 
we compare them with hydrodynamical simulations and with hybrid models run on 
N-body codes \citep[see e.g.][for an extensive discussion]{V17a}.

A few preliminary considerations are necessary.
Large scale numerical simulations are often designed to describe 
``average'' regions of the Universe, while our model is aimed at simulating a 
highly biased region, within the virial radius of a $10^{13} \, M_\odot$ DM halo 
at $z \sim 6.4$.
Small-scale simulations (box sizes up to few Mpc) are able to capture the 
scales and typical halo masses at which physical processes are important 
for early BH seeding or Pop~III star formation (such as inhomogeneous metal 
enrichment and the clustering properties of radiation 
sources). However, their predicted seed BHs are not
{directly} connected with the assembly of the first 
SMBHs at $z > 6$ \citep[see, e.g.][]{Wise12, Agarwal12, Agarwal14, 
Crosby13, Griffen16, Habouzit16hydro, Habouzit17}.
{These small-scale simulations produce enough seeds to match the number density of 
SMBHs at $z=6$. It thus becomes a matter of how many of these seeds are able to form 
in galaxy progenitors of SMBH hosts at $z=6$, and how fast these seed BHs are able to grow.}
On the opposite side, simulations with sufficiently large volumes to capture 
the formation of more massive DM halos (and SMBHs at $z\sim 6$) do not have 
the mass resolution required to follow the formation of the first stars in 
mini-halos \citep[e.g.][]{Khandai12, Feng2014, DiMatteo16, Simha17, Rong17}.
Even the most recent, high-resolution, zoom-in hydrodynamical simulations
which follow the growth of SMBHs at $z\sim 6$ \citep[e.g.][]{Sijacki09, Barai17},
do not have the adequate physical prescriptions required to describe the role
of Pop~III star formation and of different BH seeds in the final SMBH
assembly. 

The effects of metal enrichment and chemical feedback are reflected both in 
the Pop~III star formation rate density (SFRD) and in the mass distribution 
of Pop~III star forming halos. 
{ 
With the above caveat in mind, we have compared our results with the 
recent \textsc{enzo} Renaissance simulation presented by \citet{Xu13, Xu16a, Xu16b}
(``rare peak'' case) and with the Gadget-$2$ simulation by \citet{Chon16}.
We predict a Pop~III SFRD that is about $40-100$ times higher than that in
the ``rare peak'' simulation by \citet{Xu13, Xu16b}, and closer to the level 
obtained by \citet{Chon16} (assuming $h=0.73$, see their figure 2). 
If compared with the ``rare peak'' simulation presented by \citet{Xu13, Xu16b}, 
the higher SFRD that we find in our model reflects the conditions in the highly 
biased region that we are considering. 
If we estimate the over-density of our comoving volume ($50$ Mpc$^3$)  
following the definition of \citet{Xu16a}, 
$< \delta > = < \rho >/(\Omega_{\rm M}\rho_{\rm cr}) \, -1$, 
we find $< \delta > \sim 5$, a factor of $\sim 7$ higher than that of 
the ``rare peak'' simulation \citep[$< \delta > \sim 0.68$, see e.g.][]{Xu16a}.
}

In addition, we find that Pop~III stars are 
preferentially hosted in minihalos 
($6 < {\rm log}(M_{\rm halo}/M_\odot) \gtrsim 7.5$) in the redshift range 
$20\leq z < 24$ and in Lyman$-\alpha$ cooling systems 
($7.5 < {\rm log}(M_{\rm halo}/M_\odot) < 8.5$) in the redshift range $15 < z < 20$.
The recent Renaissance simulations presented by \citet{Xu13, Xu16a, Xu16b} show a 
similar mass range for halos hosting Pop~III stars, independently of the density 
of the simulated region\footnote{The Renaissance simulation has 
targeted three different subvolumes, with comoving sizes of $133.6$, $220.5$, and $220.5$~Mpc$^3$,
designated as ``Rare peak'', ``Normal'' and ``Void'' regions because they have a mean density
that is higher, comparable and lower than the cosmic average, respectively
 \citep[e.g.][]{Xu13,Oshea15,Xu16a}. In all cases, halos with mass up to few $10^9 \, M_\odot$ are formed.}.
Moreover, the fraction of halos containing Pop~III stars drops 
to zero at halo masses $<7\times 10^6 (10^7) \, M_\odot$ in their normal (high-density)
regions \citep[][]{Oshea15}. 

In much smaller boxes, simulations show that DM halos with masses
$<2\times 10^6 \, M_\odot$ do not host Pop~III stars \citep[see e.g.][]{Wise12}. 
We find a similar lower limit in V16.
In addition, \citet{Wise12} suggest that a single pair-instability supernova is sufficient 
to enrich its host halo to a metallicity level as high as $10^{-3} \, Z_\odot$, driving the 
transition to a second generation of stars in the halo already at $z>15$, similarly to what 
we find in our biased region.

It is worth noting that, while we find similar trends in the typical mass range of Pop~III star forming halos, 
we can not compare the absolute number of halos per mas bin predicted by our model with that obtained 
by \citet{Xu13, Xu16a}. As it appears evident by comparing the results of the two ENZO simulations
presented by those authors, a different number of halos (total and hosting Pop~III stars) is 
predicted in different simulated regions (high- vs low-density), even in boxes with comparable volumes.

For what concerns the effect of radiative feedback (LW radiation) on DCBHs 
formation, we have compared the evolution of the LW flux at which halos are 
exposed to with the results found by studies based on hybrid models run on 
N-body simulations in \citet{V17a} (see Fig.~3). We found that in the highly 
biased volume that we simulate, the flux level that we predict for the 
background is comparable to the maximum local flux level found by 
\citet{Agarwal12}. 
{To investigate the impact of the lack of spatial information on our DCBH seeding prescription we can derive the maximum local $J_{\rm LW}$ from our Pop III and Pop II galaxies at  fixed (physical) distances. For an escape fraction of 1, we find that the $J_{\rm LW}$ fluctuations are less important than the global background for distance scales $\geq 7$ kpc. Conversely, the local $J_{\rm LW}$ could be significantly (at least an order of magnitude) higher than our global LW background within $\sim 1$ kpc from the emitting source. \citet{Chon16} find that 2 out of 42 DC host halo candidates (i.e. $5\%$) in their simulations successfully form a DCBH when illuminated by a close-by source within few kpc ($1.27 h^{-1}$ and $6.29 h^{-1}$ kpc). We find a similar (average) fraction of atomic cooling halos hosting heavy seeds.} 

Finally, we recall that although there is a general consensus between models 
on the fundamental role of the LW radiation and of metal pollution (both in- 
and ex-situ) in setting the environment where DCBHs form, the value of the 
critical LW radiation intensity required is still highly uncertain, and there 
is a large spread in the number density of DCBHs (or DCBH host candidates) 
derived in different models \citep[see e.g.][]{Agarwal12, Agarwal14, Dijkstra14, Habouzit16hydro}.
Yet, with a critical LW threshold $J_{\rm cr}=300$, 
we predict an average DCBHs occurrence ratio of $5\%$, in agreement with recent numerical 
simulations by \citet{Chon16}.
\newline
\noindent In light of the discussion presented above, we believe that our approach is 
well motivated for the purpose of our investigation, and that it leads to robust results when 
compared with independent studies based on different techniques.

\section{Conclusions} 
\label {sec:conclusions}

In this work we have analyzed the evolution of heavy and light BH seeds, progenitors of a $z>6$ SMBH,
and their host galaxies in a cosmological context, as predicted in V16.
The aim of this study is to characterize the properties of the birth/growth environment
of these two different BH seeds. 

Our main findings are summarized as follows:

\begin{itemize}
  \item On average, light and heavy BH seeds form within a similar redshift range, 
        $15<z_{\rm form}<18$, in DM halos having comparable average mass, $\sim 5\times 10^7 \, M_\odot$.
        Even after their formation, they continue to reside in halos with similar masses 
        ($10^8-10^9 \, M_\odot$), as long as the systems evolve in isolation.

  \item About $80~(98) \%$ of light (heavy) seed hosts are found to evolve in isolation 
        (i.e. no minor or major mergers with other halos) down to $z\sim 10$. 
        Only a small number of galaxies hosting growing light (heavy) BH seeds remain 
        isolated for longer than $200~(400)$ Myr.

  \item At $z\geq 10$, galaxies hosting heavy BH seeds are characterized by a factor
        of 5-10 smaller stellar mass and metallicity than the ones hosting light seeds. 
        In addition, heavy BH seeds are accreting gas more efficiently (Eddington ratio 
        close to 1) than their lighter counterparts. 
        As a result of the efficient growth and feedback, the host galaxies of heavy BH seeds 
        experience a less intense star formation activity. Their star formation rate is about
        two orders of magnitude lower than in galaxies that host light BH seeds.

  \item At $z<10$ the fraction of isolated systems dramatically decreases to less than $2~(20)\%$, 
        for heavy (light) seed hosts, as merger events occurs. 
        The differences in the galaxy properties of different systems are progressively erased so that 
        any trace of the BH origin is lost.
\end{itemize}

We conclude that the probability to disentangle the origin of BH progenitors requires to target 
these systems at $z>10$, when their own properties and the properties of their host galaxies still
reflect/trace the conditions at BH seed formation.

The properties inspected here have a fundamental role in shaping the spectral energy distribution 
(SED) of accreting BHs and galaxies.
Indeed, the results of the statistical analysis presented here,
will be used in a companion paper (Valiante et al. 2017c) to characterize the luminosity
and colors of BH progenitors and explore the prospects of discriminating the origin of BH seeds.

\section*{Acknowledgments}
The research leading to these results has received funding from the European Research Council 
under the European Union's Seventh Framework Programme (FP/2007-2013) / ERC Grant Agreement n. 306476. 
We thank the anonymous Referee for her/his careful reading and useful comments.
We thank Marta Volonteri, Emanuel Giallongo, Andrea Grazian, Laura Pentericci and Elisabeta Lusso for useful discussions.

\bibliography{biblioReview}

\bsp
\label{lastpage}
\end{document}